\documentclass[a4paper,11pt]{article}
\usepackage[utf8]{inputenc}
\usepackage{graphicx}	
\usepackage{amsmath}	
\usepackage{amssymb}	
\usepackage{hyperref}
\usepackage{footnote}
\usepackage{natbib}
\usepackage{lineno}
\title{Discovery of massive star formation quenching by nonthermal effects in the center of NGC~1097}
\author{
F.~S. Tabatabaei$^{\star}$, 
P. Minguez,
M. A. Prieto,
J. A. Fern\'andez-Ontiveros
\\
\small Instituto de Astrof\'{i}sica de Canarias, V\'{i}a L\'{a}ctea S/N, E-38205 La Laguna, Spain\\
\small Departamento de Astrof\'{i}sica, Universidad de La Laguna, E-38206 La Laguna, Spain\\
}

\begin{document}

\maketitle

\begin{abstract}
{\bf Observations show that massive star formation quenches first at centers of galaxies. To understand quenching mechanisms, we investigate the thermal and nonthermal energy balance in the central kpc of NGC\,1097-- a prototypical galaxy undergoing quenching-- and present a systematic study of the nuclear star formation efficiency and its dependencies. This region is dominated by the nonthermal pressure from the magnetic field, cosmic rays, and turbulence. A comparison of the mass-to-magnetic flux ratio of the molecular clouds shows that most of them are magnetically critical or supported against gravitational collapse needed to form cores of massive stars. Moreover, the star formation efficiency of the clouds drops with the magnetic field strength. Such an anti-correlation holds with neither the turbulent nor the thermal pressure. Hence,  a progressive built up of the magnetic field results in high-mass stars forming inefficiently, and it may be the cause of the low-mass stellar population in the bulges of galaxies.  } 
\end{abstract}

Cosmological studies show the importance of feedback in evolution and quenching of star formation in the Universe. Caused by star formation (SF) or active galactic nuclei (AGN), feedback and its nature cannot be understood neglecting the environment where SF {and/or} AGN interact with, i.e., the interstellar medium (ISM). Simulations have been mostly based on the thermal feedback, i.e., thermal radiation pressures and thermal winds from supernovae (SNe) or AGN [1,2]. However, the nonthermal components of the ISM; magnetic fields, cosmic rays, and turbulence {(with SNe {and} AGN as one of their production sources [3,4,5])}, can be more important energetically than the thermal gas  in the ISM [6,7,8].  This motivates precise separations of the thermal and nonthermal ISM as the first step in dissecting the {origin of the feedback [9,10] and unveiling factors controlling the formation of massive stars}.

Molecular clouds, as the stellar nurseries, evolve in the ISM due to dynamical forces {caused} by gravity, turbulence, and magnetic fields [11,12,13,14] as well as stellar feedback [15,16], whose relative importance could possibly change depending on the location of clouds in a galaxy. This is actually the key point addressing where quenching begins in galaxies. Observations show that almost all quenched galaxies have a prominent bulge and a super massive black hole (SMBH) [17] and that star formation quenches first in galaxy centers [18]. This suggests detailed studies of {the ISM in centers of galaxies undergoing quenching}. Nearby galaxies hosting the most powerful astrophysical phenomena such as AGNs and starbursts are ideally suited to catch in the act of quenching of massive star formation. Hosting a Seyfert 1/LINER nucleus [19], NGC~1097 (D=\,14.5\,Mpc; 1''= 70\,pc, [20]) serves as a perfect prototype:  {With a color of [B-V]=0.85 [21], NGC~1097 belongs to the `green valley' population of galaxies showing an ongoing quenching [22,23].    }
It has yet a prominent star forming nuclear ring that has been the interest of {a number of} studies addressing its star formation rate, gas, and dust (e.g., [24,25,26]). The ring's magnetic field is the strongest {in a nearby galaxy} measured using full polarization radio continuum observations [27]. 
Taking advantage of available {\it Very Large Array (VLA)} and {\it Submillimeter Array (SMA)} observations combined with {\it Hubble Space Telescope (HST)} observations, we characterize the thermal and nonthermal ISM, investigate the ISM energy balance, and present a systematic study of the star formation efficiency and its dependencies at the center of NGC\,1097.  With a star formation rate of SFR$\simeq\,2 M_{\odot}$\,yr$^{-1}$ [24], the NGC\,1097's nuclear ring is producing massive stars less efficiently than many other starburst rings [28]. Understanding what reduces this efficiency in a galaxy experiencing quenching further provides important hints on the formation of low-mass stars in the central galaxies observed [17].

\begin{figure}
\begin{center}
\resizebox{10cm}{!}{\includegraphics*{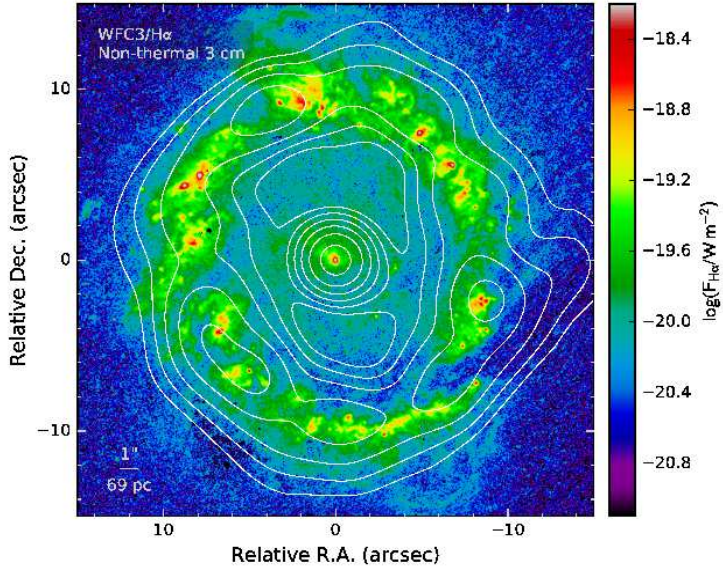}}
\caption[]{{\bf The nonthermal synchrotron emission from the central kpc of NGC~1097.} The color map shows the {\it observed} HST-WFC3 H$\alpha$ emission. After de-reddening and correcting for contamination by the [\textsc{N\,ii}]-line emission, the H$\alpha$ map  was used to trace the thermal radio (free-free) emission. Subtracting the thermal radio emission from the observed VLA radio continuum data at 3.6\,cm [27], the nonthermal synchrotron emission was extracted (contours). The contour levels are 0.25, 0.4, 0.6, 1.0, 1.6, 2.5, and 4\,mJy per 2'' beam, respectively. The H$\alpha$ map was convolved to the resolution of the radio data before separating the thermal and nonthermal emission.} 
\label{fig:tnt}
\end{center}
\end{figure}

\section{Results}

{The radio continuum (RC) emission at cm-waves is a dust-unbiased tracer of the energetic ISM components, the magnetic fields and cosmic ray electrons (CREs). It also traces the thermal ionized gas emitting the free-free emission in the radio. Separating the two RC components, the nonthermal synchrotron emission from CREs (a power-law radiation, $S_{\rm nt}\sim\,\nu^{-\alpha_{\rm nt}}$) and the optically thin thermal free-free emission from thermal electrons ($S_{\rm ff}\sim\,\nu^{-0.1}$), is the first step characterizing the ISM. We mapped the thermal and nonthermal RC emission from the central kpc region of NGC\,1097 at 3.6\,cm using the most precise separation method available (Sect.~3, Fig.~1).}  The thermal fraction ($f_{\rm th}=S_{\rm ff}/S_{\rm RC}$, with $S_{\rm RC}=S_{\rm ff}\,+\,S_{\rm nt}$) changes between 24\% and 64\% in the nuclear ring with an average of  40\%\,$\pm$\,3\%.
The southern ring is mostly nonthermal in nature: The thermal fraction is $\sim$\,35\% in the southern clumps, while $\sim$\,58\% in the northern ring on average. 
{The very central region ($R<$\,0.1\,kpc)} also appears as a nonthermal source ($f_{\rm th}<$20\%) possibly due to an AGN jet [29].

\begin{figure}
\begin{center}
\resizebox{12cm}{!}{\includegraphics*{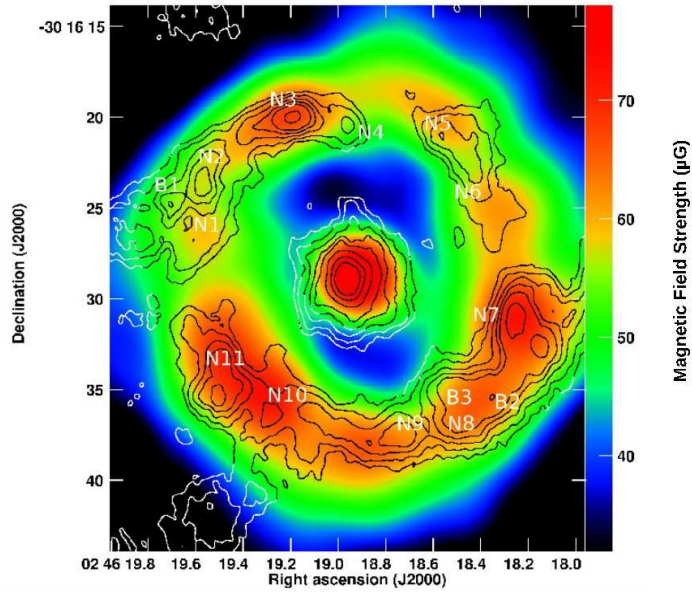}}
\caption[]{{\bf Magnetic field and molecular clouds.} The equipartition magnetic field strength B mapped in   the central kpc of NGC~1097 is by a factor of $\gtrsim$\,3 stronger than average B in normal star forming galaxies [9]. The origin of such a strong field is an open question (see Fig.~3). It is relatively stronger in denser molecular clouds traced in CO(2-1) line emission (contours). The contour levels show 3$\sigma$, 4$\sigma$, 6$\sigma$, 8$\sigma$, 10$\sigma$, 13$\sigma$, 17$\sigma$, 22\,$\sigma$  with 1$\sigma$\,=\,2.3\,Jy\,kms\,$^{-1}$ per $1''.5\times 1''.0$ synthesized beam of the SMA observations [24]. On the map, we also label the GMAs listed in Table~1.}
\label{fig:B}
\end{center}
\end{figure}

{The thermal and nonthermal RC maps provide measures of the ISM physical parameters, the thermal electron density of the warm ionized gas (WIM) and the magnetic field strength, respectively (Sect.~3).}  
In the nuclear ring of NGC\,1097, the thermal emission leads to a WIM volume-averaged electron density of $<n_e>\,=2.0\,\pm\,0.1$\,cm$^{-3}$. {This translates to a {\it dense ionized cloud} density of $n_e~\simeq40$\,cm$^{-3}$ (Sect.~3) which agrees with the dust-unbiased far-infrared line ratio measurements in the center of NGC\,1097 [30], indicating the reasonable reddening laws used in our method.} $<n_e>$ is larger than that found in the Galaxy ($\sim$\,0.02\,cm$^{-3}$ [31]) by 2 orders of magnitude.  %
Table~\ref{tab:list} lists $<n_e>$ as measured {over the giant molecular cloud associations (GMAs)} in the ring.  

The map of the equipartition magnetic field strength B shows that the ring is strongly magnetic with B changing between 50 and 80\,$\mu$G and a mean of $62\pm2\,\mu$G (Fig.~\ref{fig:B}). This is about 13\% higher than the  mean value of $\sim55\,\mu$G given by [27]. %
In between the ring and the nucleus, B reduces to $40\,\mu$G on the average. The field strength is the largest at the nucleus with  B=\,$90\pm\,3\mu$G. Subtracting its ordered part 
following [32], B appears to be mostly random  in the nuclear ring  (B$_{\rm ord}$/B$\sim$\,10\%).

\begin{table}[!ht]
 \centering
 \small
\caption{{\bf Thermal and nonthermal properties of molecular cloud associations.} We list the coordinates and the physical properties of the GMAs used and/or obtained-- $\delta$R.A. and $\delta$Dec.: offset positions from the galaxy center ($\alpha_{2000}=02^{\rm h}\,46^{\rm m}\,18^{\rm s}.96$ and $\delta_{2000}=-30^{\circ}\,16'\,28''.897$), $d$: diameter, $\Sigma_{\rm H_2}$: molecular gas mass surface density adopted for X$_{\rm CO}=1\times 10^{20}$\,cm$^{-2}$\,(K\,km\,s$^{-1}$)$^{-1}$,  $\sigma_v=\delta$V$_{\rm int}/\sqrt{8\,{\rm ln2}}$ with $\delta$V$_{\rm int}$ the intrinsic CO(2-1) line width, B: {\it measured} magnetic field strength (B$_{\rm tot}$=\,1.3\,B, and B$_{\rm tot}^2$=1.5\,B$^2$),  $\mu$: mass-to-magnetic flux ratio in units of $\mu_0=(2\pi\,G^{1/2})^{-1}$, $E_{\rm k}/E_{\rm b}$: ratio of the kinetic energy density to magnetic energy density, $<n_e>$: volume averaged electron density, $\beta$: ratio of the thermal energy density to magnetic energy density ($={\rm E_{\rm th}/E_{\rm b}}$), and $\Sigma_{\rm SFR}$: star formation rate surface density. The symbols $\star$ and $\maltese$ refer to {\it this work} and [24], respectively.
The uncertainty is 0.01\,g\,cm$^{-2}$ in $\Sigma_{\rm H_2}$,  and 0.06\,cm$^{-3}$ in $<n_e>$. Both statistical (1\,$\sigma$) and systematic calibration errors were propagated to estimate the total uncertainties in B, $\mu$, $E_k/E_b$, $<n_e>$, and $\beta$. } 
\begin{tabular}{ l l l l l l l l l l l l } 
\hline
\hline
 & & & & & & & & & & & \\
Cloud  &  $\delta$R.A.$^{\maltese}$   & $\delta$Dec.$^{\maltese}$ & $d^{\maltese}$ & $\Sigma_{\rm H_2}^{\maltese}$ & \,\,\,$\sigma_v^{\maltese}$& \,\,\,\, B$^{\star}$ &\,\,\,\, \,\,\,\,\, $\mu^{\star}$& $E_k^{\star}/E_b$&$<n_e>^{\star}$ & \,\,\,\,$\beta^{\star}$ &\,\, $\Sigma_{\rm SFR}^{\maltese}$\\ 
Name & \,\,\,\,\,('') & \,\,\,('')   & ('')   &(g & (km\,s$^{-1}$)&  \,\,\,\,($\mu$G) & \,\,\,\, \,\,\,\,($\mu_{0}$)&  &\,(cm$^{-3}$) & $\times10^{-2}$ & ($M_{\odot}$\\ 
 & & & &\,cm$^{-2}$) & & & & & & & yr$^{-1}$\,kpc$^{-2}$)\\
\hline \hline
N1      & \,+9.0  & \,+1.6   &   2.9  &0.09 & 17.8$\pm$1.7& $63\pm 3$ & 0.84$\pm$0.12 & 0.6$\pm$0.1 & 3.5 & 3.2$\pm$0.1  & 2.74$\pm$0.25\\
N2      & \,+8.6 & \,+4.8  &     3.7     &0.14 & 25.9$\pm$1.7 & $62\pm 3$& 1.43$\pm$0.13 & 2.3$\pm$0.2 &3.7    & 3.7$\pm$0.1      &3.18$\pm$0.18 \\
N3      & \,+3.6 & \,+8.4   &    3.5     &0.20 & 27.2$\pm$0.4& $71\pm 2$ & 1.73$\pm$0.11 & 2.6$\pm$0.2 & 3.3& 2.6$\pm$0.1 & 1.81$\pm$0.10\\
N4      &\,+0.6  & \,+8.2   &     2.2       &0.09 & 23.7$\pm$1.7& $61\pm 3$ & 0.95$\pm$0.12 & 1.3$\pm$0.2 &4.1& 3.9$\pm$0.2& 4.06$\pm$0.28\\
N5      &\, -4.8  & \,+7.8 &       2.9         &0.11 & 13.2$\pm$0.4 & $61\pm 3$ & 1.12$\pm$0.15 & 0.5$\pm$0.1 &3.5 & 3.5$\pm$0.1& 3.41$\pm$0.17\\
N6      & \,-6.4 &  \,+5.6  &      2.9          &0.10 & 15.7$\pm$0.8& $58 \pm 3$ & 1.09$\pm$0.13  & 0.7$\pm$0.1 &3.4&  3.8$\pm$0.1&3.60$\pm$0.17\\
N7       & \,-8.6 & \,-2.6 &        3.4         &0.21 & 22.1$\pm$1.3& $70 \pm 2$ & 1.84$\pm$0.11  & 1.8$\pm$0.1 &2.8&  2.3$\pm$0.1 &4.07$\pm$0.34\\
N8     &   \,-5.4 & \,-8.4 &        2.5          &0.15 & 21.6$\pm$0.8 & $67\pm 2$& 1.37$\pm$0.12 &  1.4$\pm$0.1 &2.7& 2.4$\pm$0.1& 1.91$\pm$0.19\\
N9     &  \,-2.8 &\,-9.2   &       3.7              & 0.13 & 24.2$\pm$1.3 & $69\pm 2$& 1.10$\pm$0.11  &  1.4$\pm$0.1 & 3.0& 2.4$\pm$0.1& 1.69$\pm$0.16\\
N10     & +4.8 & \,-7.8  &         2.8        &0.10 & 15.7$\pm$0.8 & $72 \pm 2$& 0.85$\pm$0.10 &   0.4$\pm$0.1 &2.6& 1.9$\pm$0.1 &1.32$\pm$0.14\\
N11     &  +7.8 & \,-6.6  &         4.1       &0.14 & 20.8$\pm$1.7 & $73 \pm2 $& 1.22$\pm$0.10 &  1.0$\pm$0.1 &2.6& 1.9$\pm$0.1& 0.87$\pm$0.10 \\
Nu     &   \,\,0.0 & \,\,0.0&      ...        &0.26 & 41.6$\pm$8.5 & $90\pm 3$&1.81$\pm$0.09 & 5.1$\pm$1.1 & 2.7& 1.4$\pm$0.1& \,\,\,\,\,\,\,\,..... \\
B1     &   10.2 & \,+3.6  &          3.1         &0.14 & 46.3$\pm$4.7 &$62 \pm 2$ & 1.43$\pm$0.13 & 7.1$\pm$1.0 & 0.9 &  3.2$\pm$0.1&  1.16$\pm$0.09\\
B2    &  \,-7.4 &  \,-6.8    &       3.3        & 0.18 & 40.8$\pm$2.1 & $71 \pm 2 $ & 1.57$\pm$0.11 &   5.3$\pm$0.5 & 2.5 & 2.0$\pm$0.1& 1.18$\pm$0.11\\
B3  &   \,-5.4  & \,-6.6     &       3.0        & 0.16 & 38.2$\pm$3.4 & $63 \pm 2$ &  1.59$\pm$0.12  & 5.3$\pm$0.6 & 2.0 & 2.1$\pm$0.1 & 0.58$\pm$0.08\\
\hline
Ring  & \,\,.....&\,\,.....& ....&0.12 & \,\,\,\,\,\,\,20 & \,\,\,$62$ & \,\,\,\,\,\,\,..... & \,\,\,\,\,\,\,0.4& 2& \,\,\,\,\,\,\,3&\,\,\,\,\,\,\,\,2 \\
\hline \hline
\label{tab:list}
\end{tabular}
\end{table}

\begin{figure*}
\begin{center}
\resizebox{10cm}{!}{\includegraphics*{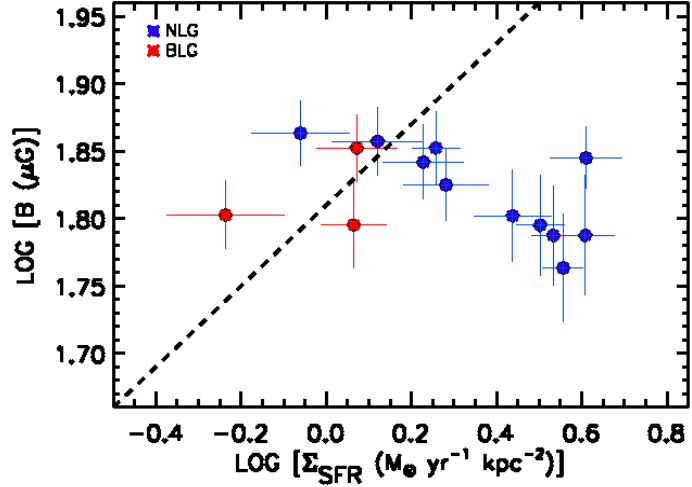}}
\caption[]{ {\bf Magnetic field and star formation activity in GMAs.} The equipartition magnetic field strength B does not scale with the star formation rate surface density $\Sigma_{\rm SFR}$ unlike in galaxy disks (e.g., [32]). The dashed line shows the theoretical proportionality of B$\sim\,  \Sigma_{\rm SFR}^{0.3}$  as expected for amplification of B in star forming regions [34]. The magnetic field coupled with gas is likely enhanced due to the gas flow to the center and as a result of local shear and shocks [35].   The blue and red points show the narrow-  and broad-line GMAs, respectively. The errors in B show both statistical and systematic uncertainties. The horizontal error bars show the uncertainty in $\Sigma_{\rm SFR}$ given by [24].  }
\label{fig:SFR-B}
\end{center}
\end{figure*}

The magnetic field in the center of NGC\,1097 is by a factor of $\gtrsim$\,3 stronger than its average in normal star forming galaxies (a median of $13.5\pm 5.5$\,$\mu$G in the KINGFISHER sample, [9]). The origin of such a strong field is an open question. One general theory to explain the galactic magnetic field is the small-scale dynamo converting kinetic energy of turbulence to magnetic energy [33]. Using a SNe-driven dynamo model with B$_{\rm ran}$/B$_{\rm ord}=2-3$ [3] and assuming equipartition between magnetic field and cosmic rays, [34] obtained a relation between B and the star formation rate surface density ($\Sigma_{\rm SFR}$), B$\sim\,\Sigma_{\rm SFR}^{0.3}$. A similar relation between B and the global SFR has been observed in nearby galaxies (e.g. [9]).  However, according to this relation and using the galaxy median values as reference (e.g., B$\sim$13.5\,$\mu$G, SFR$\sim 1 M_{\odot}$\,yr$^{-1}$ from the KINGFISHER sample [9]), {the SFR must be $\sim$75 times higher than its actual value of 2\,$M_{\odot}$\,yr$^{-1}$ [24]. More surprisingly, B is found to be almost anti-correlated with $\Sigma_{\rm SFR}$ in the GMAs (Fig.~\ref{fig:SFR-B}) and hence it cannot be attributed to the star formation feedback. The magnetic field must have been enhanced due to a strong gas compression {and/or} local shear [35] in the galaxy center}. Models should be modified for such an extreme condition with B$_{\rm ran}$/B$_{\rm ord}\simeq\,9$ (in the ring) and accounting for turbulence inducing sources such as dynamical pressure {and} shocks due to gas flow to the center and amplification due to a {\it gravity-driven turbulence} [36]. In spite of its origin, the presence of such a strong magnetic field could possibly change the ISM energy balance and {further inserts  {\it 'a nonthermal ISM feedback'} controlling the molecular  clouds and formation of stars}. These are further discussed as follows. 

\section{Discussion}
\subsection{ISM Energy Balance in The Ring's Volume}
{The ISM physical state and structure formation is controlled by a balance between the thermal and nonthermal energy densities. } The energy content of the ISM is set by the thermal gas, turbulent motions, and the nonthermal pressures inserted by the magnetic field and cosmic rays.  
The energy density of the magnetic field is E$_{\rm b}$=\,B$^2 /8 \pi$. In case of equipartition between the CREs and magnetic energy densities, the total nonthermal energy density due to both CREs and B is E$_{\rm nt}$=\,2\,E$_{\rm b}$= B$^2 /4 \pi$.  {In the nuclear ring of NGC\,1097, we find a mean nonthermal energy density of}  E$_{\rm nt}\simeq\,5\times\,10^{-10}$ erg\,cm$^{-3}$. 

The thermal energy density of the ionized gas, $\frac{3}{2} \langle n_e \rangle kT_e$, is estimated for {WIM ($T_e \simeq 10^4$\,K) and $\langle n_e \rangle$ determined in Sect.~1}. Assuming the pressure equilibrium between the warm and hot ionized gas with  $T_e \simeq 10^6$\,K and an electron density of $\simeq 0.01 \langle n_e \rangle$ [37], the energy density of the hot ionized gas is about the same order of magnitude as the warm ionized gas energy density. Hence, in the ring, the thermal energy density of the ionized gas is $\simeq\,8.3\times\,10^{-12}$ erg\,cm$^{-3}$ (for $<n_e>$ $\simeq$ 2\,cm$^{-3}$) and smaller in the nucleus.   
The neutral gas is mostly molecular here (the inner 1'.6 region of NGC\,1097 is HI-deficient, [38]).  The average ring gas mass of $1.7\times\,10^7\,M_{\odot}$ obtained by [24] is equivalent to a mass surface density of 1.7\,$\times\,10^{3}\,M_{\odot}$\,pc$^{-2}$ for a CO-to-H$_2$ conversion factor of X$_{\rm CO}=3\times 10^{20}$\,cm$^{-2}$\,(K\,km\,s$^{-1}$)$^{-1}$. However, [24] caution that this factor overestimates the gas mass in the Galactic center by a factor of 2-5. Our study in a similar barred spiral galaxy, NGC\,1365 [39], resulted in a  X$_{\rm CO}$ value of $1\times 10^{20}$\,cm$^{-2}$\,(K\,km\,s$^{-1}$)$^{-1}$ in the central 4\,kpc region {using which} leads to a mass surface density of $\Sigma_{\rm H_2}\simeq 0.6\,\times\,10^{3}\,M_{\odot}$\,pc$^{-2}$ in the nuclear ring of NGC\,1097. The thermal energy density of the cold neutral gas (T=20\,K, {[24]}) is hence $\simeq\,10^{-12}$ erg\,cm$^{-3}$ in the ring with a width of 300\,pc {as measured in Sect.~3}  ($\rho=\frac{\Sigma_{\rm H_2}}{300}\,M_{\odot}$\,pc$^{-3}$). The warm neutral gas with a typical temperature of $\simeq 6\,\times 10^3$\,K has roughly the same thermal energy density as the cold neutral gas, due to a $\simeq100$ times smaller density [37]. 
The total thermal energy density including the contribution of the warm, hot ionized and the cold, warm neutral gas, is hence E$_{\rm th}\simeq\,10^{-11}$ erg\,cm$^{-3}$ {in the ring volume.} 

The kinetic energy density of the gas turbulent motions, E$_{\rm k}=\frac{1}{2}\, \rho\, {\sigma_v}^2$, depends on the turbulent velocity $\sigma_v$. Using the intrinsic CO(2-1) line width $\delta$V$_{\rm int}$ given by [24], we obtain $\sigma_v=\delta$V$_{\rm int}/\sqrt{8\,{\rm ln2}}\sim\,20$\,km\,s$^{-1}$  (same $\sigma_v$ is obtained using the ALMA-HCN observations [25]) {yielding a } turbulent energy density of $2.5\times 10^{-10}$ erg\,cm$^{-3}$. 

Hence, {in the nuclear ring (and generally at $R<1$\,kpc), the ISM is controlled by the nonthermal and turbulent processes as E$_{\rm nt}\gtrsim$\,E$_{\rm k}>$\,10\,E$_{\rm th}$. This conclusion is not affected by the major sources of the uncertainty, i.e., the assumed $T_e$ for the ionized gas and the adopted X$_{\rm CO}$ factor for the molecular gas.  Fluctuations in $T_e$ ($5\times 10^3\,{\rm K}\,\lesssim\, T_e\, \lesssim\,2\times 10^{4}\,{\rm K}$) would change the WIM energy density by less than a factor of 1.6 (taking into account the dependency of $<n_e>$ on $T_e$). Any variation in $\rho$ due to the X$_{\rm CO}$ factor changes E$_{\rm th}$,  E$_{\rm k}$, and E$_{\rm nt}$ similarly (E$_{\rm nt}\propto\rho$ taking into account the magnetic field--gas coupling of B$\propto\,\rho^{0.5}$). Thus the inequality E$_{\rm nt}>$~E$_{\rm th}$ is unaffected by the assumptions used.} 

{The center of the galaxy is then occupied by} a low-$\beta$ plasma ($\beta\equiv\,$E$_{\rm th}$/E$_{\rm b}<1$) and the turbulence is supersonic as E$_{\rm k}$/E$_{\rm th}~>~1$. The supersonic motions of the turbulence {are  Alfv\'enic as the Alfv\'en  speed $V_{\rm A}={\rm B}/(4\pi\,\rho)^{1/2}$ is larger than the sound  speed $c_s= (kT_{\rm K}/2.33m_{\rm H})^{1/2}$ (with $T_{\rm K}$ the kinetic temperature and $m_{\rm H}$ the Hydrogen mass) by more than one order of magnitude}  in the nuclear ring. Hence, the ISM is in favor of the 3-D MHD models with a Mach number of $M_{\rm A}= \sqrt{3}\,\sigma_v/V_{\rm A}\sim1$.  %

\subsection{What Controls Nuclear Star Formation}
\begin{figure*}
\begin{center}
\resizebox{\hsize}{!}{\includegraphics*{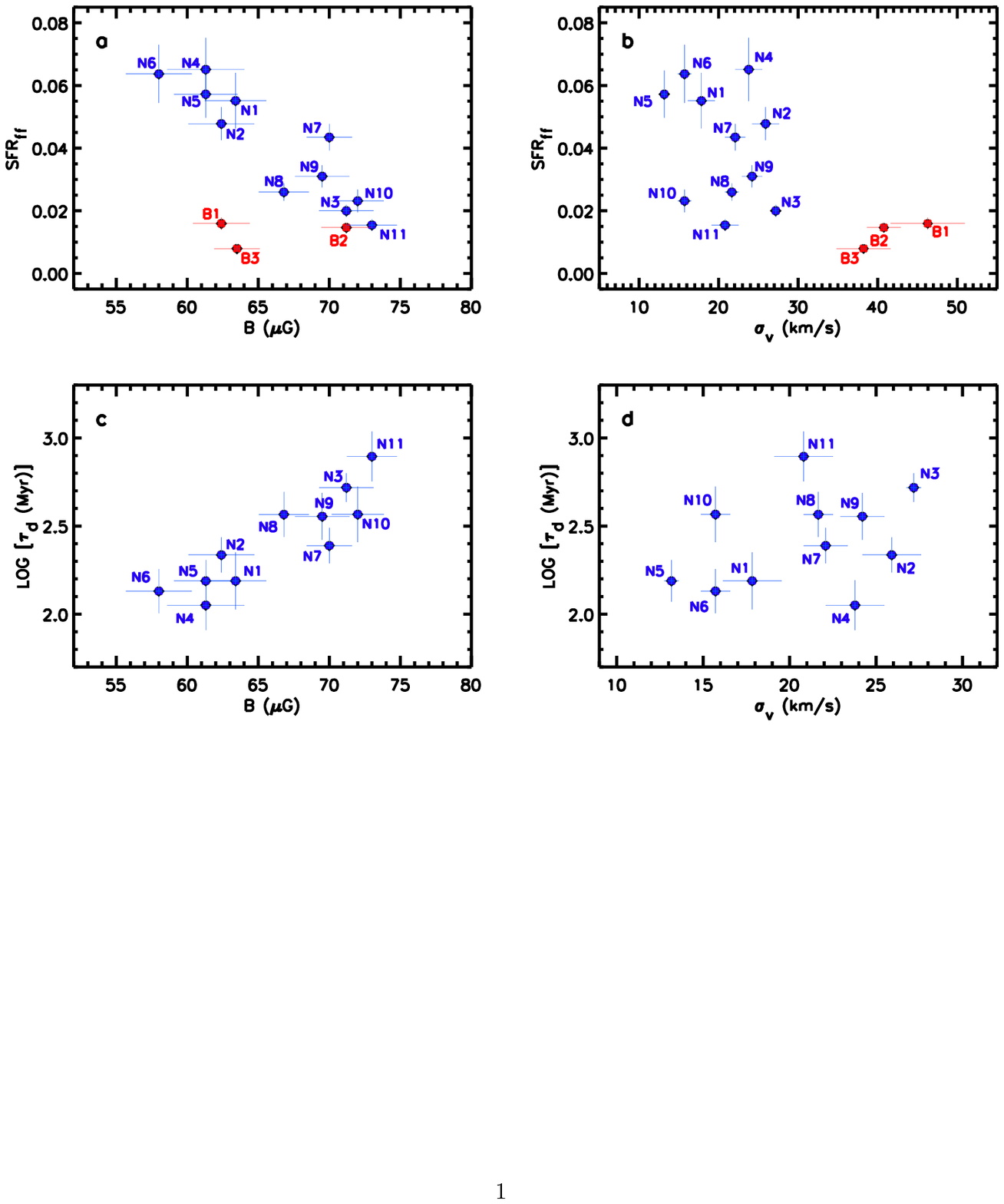}}
\caption[]{ {\bf Clouds with stronger magnetic fields are less efficient in forming massive stars.} The massive star formation rate per free-fall, SFR$_{\rm ff}$, of the GMAs decreases with the magnetic field strength B ({\it a}) while it is uncorrelated with the turbulent velocity $\sigma_v$ ({\it b}).  The blue and red points show the narrow-  and broad-line GMAs, respectively. Strong non-circular motions/shocks in the broad-line GMAs can act as additional cause of the low SFR$_{\rm ff}$ in these clouds.  A longer time is needed to form massive stars in a stronger magnetic field, as the molecular gas depletion timescale $\tau_{\rm d}$ increases with B  ({\it c}). We also note that no correlation holds between $\tau_{\rm d}$ and $\sigma_v$ ({\it d}). The vertical error bars  show the uncertainties in  SFR$_{\rm ff}$ ({\it a} and {\it b})  and $\tau_{\rm d}$ ({\it c} and {\it d}) calculated by propagating the errors in  $\Sigma_{\rm SFR}$, $\Sigma_{\rm H_2}$, (and $M_{\rm H_2}$ for SFR$_{\rm ff}$) given by [24]. The horizontal error bars show the uncertainties in B ({\it a} and {\it c}) representing both statistical and systematic uncertainties and $\sigma_v$ ({\it b} and {\it d}) given by [24].}
\label{fig:SFE}
\end{center}
\end{figure*}

An interesting feature in Fig.~\ref{fig:B} is the clumpiness of the magnetic field in the nuclear ring. These clumps {globally coincide with the molecular clouds and the GMAs traced by the CO(2-1), and CO(3-2) line emission, particularly in the southern ring. (The Spearman rank correlation between B and $\Sigma_{\rm H_2}$ of the GMAs is $r_s=0.56\pm0.07$}.) This conveys an important message: The synchrotron emission can trace the magnetic fields coupled with not only the ionized gas but also the {neutral molecular gas down to the GMA scales. Such a correlation between the pure synchrotron emission and $\Sigma_{\rm H_2}$ had been shown before at a larger spatial scale of 0.6\,kpc in NGC\,6946 [32]. {More generally, a correlation between the observed radio continuum emission and the molecular gas emission holds in galaxies (e.g. [40])}. The nonthermal and thermal properties of the GMAs including B, $\sigma_v$, and $<n_e>$  listed in Table~1 allows uniquely a statistical study of the impact of the ISM  and the star formation feedback in cloud {and} star formation and further help evaluating the theories.  Modern MHD numerical experiments show that massive stars form in a slower mode in a stronger magnetic field [11,12]. To compare the star formation efficiency of the  GMAs, we obtained the star formation rate per free-fall SFR$_{\rm ff}$, that is a dimensionless efficiency (Sect.~3), and investigated its variations in the sample.  The SFR$_{\rm ff}$ is particularly lower in the southern ring where B and $\sigma_v$ are both higher (Fig.~\ref{fig:B} and Table~1). Do the GMAs with a lower SFR$_{\rm ff}$ have a stronger B or $\sigma_v$? Figure~\ref{fig:SFE} shows that the SFR$_{\rm ff}$ decreases with B ($r_{\rm s}=-0.93\pm0.01$ for the narrow-line GMAs), while no correlation holds with $\sigma_v$ or E$_{\rm k}$ ($r_{\rm s}=-0.24\pm0.5$). Moreover, no decreasing trend is found between the SFR$_{\rm ff}$ and the thermal pressure (E$_{\rm th}$) tracing the star formation feedback. 
The tight correlation between the molecular gas depletion timescale $\tau_{\rm d}$ and B (Fig.~\ref{fig:SFE}-c) shows a slower formation of the massive stars in a stronger magnetic fields. {This provides a robust observational evidence} for the theoretical predictions [11, 13] in a sample of clouds in a galaxy center. 

It is worth noting that the trends observed demonstrate a negative feedback from the nonthermal ISM regardless of the assumptions made to estimate B because similar correlations (over a wider dynamical range) hold with the synchrotron intensity $S_{\rm nt}$ itself.  If no CRE--B equipartition holds, B obtained will only represent a scaled version of $S_{\rm nt}$ (as B$\propto S_{\rm nt}^{\frac{1}{4}}$ was used) and Figs.~\ref{fig:SFE}-a,c show a combined effect from CREs and B, i.e, from the nonthermal ISM. We also stress that $\gamma$-ray observations confirm the validity of the CRE--B equipartition for $\Sigma_{\rm SFR}<100\,M_{\odot}$yr$^{-1}$\,kpc$^{-2}$ [41], and hence for the nuclear ring of NGC\,1097 (with $\Sigma_{\rm SFR}\simeq\,2\,M_{\odot}$yr$^{-1}$\,kpc$^{-2}$).   } 

{Investigating further the role of the magnetic field in evolution of the GMAs, we determine the mass-to-magnetic flux ratio M/$\phi$ ($\equiv\mu$ expressed in units of the critical value, $\mu_0=(2\pi\,G^{1/2})^{-1}$ [42]) as a standard parameter assessing} to what extent a static magnetic field can support a cloud against gravitational collapse. Clouds will not collapse due to self-gravity when $\mu<1$ (see e.g. [13, 43]).
By definition $\mu$ is given by the column density perpendicular to the sheet of matter in a flux tube, $N_{\perp}$ and the total magnetic field, B$_{\rm tot}$, in that tube (magnetically supported cloud), $\mu= N_{\perp}/{\rm B}_{\rm tot}$. Generally, observed magnetic fields from the clouds are always smaller than the intrinsic total fields. Our synchrotron emissivity approach measures the plane-of-the sky component of B$_{\rm tot}$. Through a statistical study, [44] showed that B= 0.79\,B$_{\rm tot}$,  B$^2$= 2/3\,B$_{\rm tot}^2$, and also the observed column gas density $N_{\rm obs}=2N_{\perp}$. Hence, we use  $N_{\perp}=\Sigma_{\rm H_2}/2$, B$_{\rm tot}$=\,1.3\,B, and B$_{\rm tot}^2$=1.5\,B$^2$ calculating $\mu$, $\beta$, and E$_{\rm b}$ for the GMAs (see Table~1). 
We find that most narrow line GMAs (listed as N$i$ with $i=1,2,...,11$ in the Table) are magnetically critical as $\mu\simeq1$, and some slightly subcritical (N1, N4, and N10). The GMAs with lower densities have smaller $\mu$ and hence a larger magnetic support.  A subcritical condition is particularly evident in the low density regions in the southern ring (e.g., those surrounding the N10\,\& N11 clumps, Fig.~\ref{fig:B}). The N3 and N7 clumps which are the ring's densest and strongest in the HCN emission [25] have currently the least magnetic support in the ring.  The broad line GMAs  B1, B2, and B3 as well as the nucleus, Nu, which have ${\sigma_v}$ twice larger than the rest {of the} GMAs (we note that ${\sigma_v}$ represents not only the random dispersion but also the non-circular motions generated by shock fronts [24] which could be non-negligible for B1, B2, B3, and Nu, see e.g.\,[45]) also appear as magnetically supercritical. We stress that according to a more realistic model, in which the magnetic flux tubes are differently loaded with mass, the critical M/$\phi$ should be larger by a factor of 2 [46], based on which all the GMAs in this study fall into the subcritical picture. 
We also note that the synchrotron method can generally underestimate B in molecular clouds: Magnetic fields can be stronger by a factor of $\sim$100 based on CO polarization observations [47] {magnifying their important impact. }  

{As in the global ring case, the ISM at the GMA scale is Alfv\'enic, supersonic, and a low-$\beta$ plasma with $\beta<$0.04, lower than that in the Milky-Way clouds [13]. 
The  magnetically supported condition is continued and even enhanced in between the ring and nucleus due to its low gas density [24,25] and relatively strong magnetic field (B$\sim\,40\,\mu$G).}

{Magnetic fields supporting the molecular clouds prevent collapse of gas to densities needed to form massive stars (that is $\geq1$g\,cm$^{-2}$ [48]). Accordingly,  cloud fragmentation will continue to reach the regime for the low density gas to form many but low mass stars   [48, 14]. 
{In other words,  high gas densities needed to account for the enhanced star formation are self-regulating: High gas densities in galaxy centers lead to strong magnetic fields which ultimately limit the efficiency of massive star formation while fostering enhanced low mass star formation. This is in favor of the central low-mass stellar population and big bulges observed in quenched galaxies. 

Hence, this study unveils the role of the nonthermal ISM in quenching the massive star formation at the center of NGC\,1097- a prototypical galaxy undergoing quenching.  Observations in nearby galaxies shows that the synchrotron emission from the centers of the SMBH hosts is much stronger than that from their disks [e.g., 32, 49, 50] suggesting that a strong B is a general property of the SMBH-host galaxies. Thus, an effectively negative feedback from the nonthermal ISM could be common in quenched galaxies taking into account that almost all of them host a SMBH [17].   This prompts in-depth studies of the magnetized ISM in complete samples of galaxies being quenched and further evokes inclusion of the magnetic fields and cosmic rays in the galaxy evolution research. Such studies will be ideally possible in resolved and sensitive radio surveys of galaxies near and far with the forthcoming detectors and arrays such as the square kilometer array--SKA. }

\paragraph{Acknowledgments} We thank P.Y. Hsieh and R. Beck for providing us with the SMA-CO and the VLA-3.6cm data. F.S.T. and M.A.P. acknowledge financial supports from the Spanish Ministry of Economy and Competitiveness (MINECO) under grant numbers AYA2016-76219-P and MEC-AYA2015-53753-P, respectively.

\paragraph{Author contributions} F.S.T. conceived, designed the project, provided the thermal/nonthermal separation code, analyzed the data, wrote the paper. P.M. co-analyzed part of the data and contributed in materials. M.A.P. helped in setting up of the project. J.A.F.O. obtained the continuum-subtracted H$\alpha$ and Pa$\alpha$ maps and produced Fig.~1. MAP and JAFO commented on the manuscript and were involved in the science discussion.

\paragraph{Correspondence} and requests for materials should be addressed to FST (email: ftaba@iac.es).

\section{Methods}
\label{sec:method} %
{To separate the thermal and nonthermal components of the RC emission, we used the thermal radio tracer (TRT) technique [32,50,51],  in which one of the hydrogen recombination lines (often H$\alpha$) is used as a free-free tracer. The results of the TRT method are more reliable and precise than those of the classical method, i.e., using only the radio data and assuming a fixed $\alpha_{\rm nt}$ (as it neglects variations in the CRE energy index [32,50,51]).} The HST observations of NGC\,1097 enabled us to map the RC components using the de-reddened H$\alpha$ (or the Pa$\alpha$) emission as the thermal tracer with unprecedented {sensitivity}.  

Narrow-band images including the H$\alpha$+[\textsc{N\,ii}]$\lambda 6548, 6584$ blend and the Pa$\alpha$  ($1.87\,\mu$m) emission line were acquired with the Wide Field Camera 3 (WFC3, filter F657N) and the Near Infrared Camera and Multi-Object Spectrometer (NICMOS, filter F187N), respectively, onboard the HST. The reduced and flux-calibrated images were taken from the scientific archive and registered using seven point-like clusters in the starburst ring as reference for the alignment following [52]. The continuum was estimated using a linear interpolation between the WFC3 broad-band filters F547M and F814W for the H$\alpha$ image, and the NICMOS narrow-band filters F187N and F190N for the Pa$\alpha$ image. The contribution of the [\textsc{N\,ii}] was subtracted using the [\textsc{N\,ii}]/H$\alpha$ ratios given by [53]. The rms noise measured in dark regions of the H$\alpha$ and Pa$\alpha$ maps are $4.3\times\,10^{-20}$ and $10^{-18}$ erg\,s$^{-1}$\,cm$^{-2}$, respectively. A 5\% systematic error due to calibration and continuum subtraction was also used in both images. 

{We used the RC data at 3.6\,cm (8.45\,GHz) taken with the Very Large Array (VLA) in DnC and CnB configurations (https://www.mpifr-bonn.mpg.de/atlasmag, [27]). The total intensity map has an rms noise level of 9$\mu$Jy per 2'' beam and a systematic calibration uncertainty of 2\%.
The H$\alpha$ and Pa$\alpha$ images were normalized to the same geometry, grid size, and spatial resolution as of the RC data before the analysis. } 

The H$\alpha$ (or Pa$\alpha$) emission was de-reddened using the Balmer-to-Paschen line ratio method. The extinction at the H$\alpha$ wavelength, $A_{\rm H\alpha}$, was obtained following [54] for a Milky Way type reddening $R_v=3.1$ and an intrinsic H$\alpha$/Pa$\alpha$=\,8.62 [55]. 
The intrinsic H$\alpha$ emission was derived through $ I_{\rm H\alpha}^{\rm int}= I_{\rm H\alpha}^{\rm obs}\, 10^{0.4\,A_{\rm H\alpha}}$ and then converted to the thermal free-free emission $S_{\rm ff}$ assuming case B recombination following [32,50,51]. (Similarly, the Pa$\alpha$ emission can be used instead of the H$\alpha$ emission.)
We calculated $S_{\rm ff}$ at 3.6\,cm (4.85\,GHz) for all pixels in the observed area and constructed the nonthermal intensity $S_{\rm nt}$ map following $S_{\rm nt}=S_{\rm RC} - S_{\rm ff}$,
with $S_{\rm RC}$ the observed RC intensity at 3.6\,cm. The uncertainties were calculated by propagating the errors due to both statistical and systematic calibration errors {leading to a mean uncertainty of 4\% and 6\% in the thermal and nonthermal maps, respectively.} 

{We used the thermal RC emission to obtain the electron density $n_e$.} The thermal emission measure from an ionized gas is given by $EM = \int{ n_e^2.\,dl}=\,<n_e^2> L$, with $L$ the pathlength.  The volume-averaged electron density along the line of sight is given by $<n_e>=\sqrt{f<n_e^2>}$, with $f$ the volume filling factor describing the fluctuations in $n_e$, $f~\equiv~\frac{<n_e>}{n_e}$, that is about 5\% based on both Galactic and extra-galactic observations  [31, 56].  %

The strength of the total magnetic field B is derived using the nonthermal synchrotron intensity $S_{\rm nt}$ and assuming equipartition between the energy densities of the magnetic field and cosmic rays
($\varepsilon_{CR}=\varepsilon_{\rm B}={\rm B}^2/8\pi$):
%
${\rm B}~=~C(\alpha_{\rm nt}, K, L) \big[S_{\rm nt} \big]^{\frac{1}{\alpha_{\rm nt+3}}}$,
where $C$ is a function of $\alpha_{\rm nt}$, $K$ the ratio between the number densities of cosmic ray protons and electrons ($K\simeq\,100$, [57]). 
B is obtained assuming that the magnetic field is parallel to the plane of the galaxy (inclination of $i=46^{\circ}$ and position angle of the major axis of PA=-45$^{\circ}$, [38]) and $\alpha_{\rm nt}=1$. {Assuming that B is fully isotropic results in an 8\% reduction in the magnetic field strength in every location}.  As discussed earlier, $\alpha_{\rm nt}$ should be flatter in the  star forming regions. We have shown that $\alpha_{\rm nt}\simeq\,0.6$ in star forming complexes in nearby galaxies [32,51], using which leads to a maximum 10\% increase in B in the nuclear star forming clumps of NGC~1097. 
Both B and $<n_e>$ were calculated for $L=300\pm35$\,pc that is the projected width of the nuclear ring.  $L$ is taken as the FWHM of a Gaussian fit at the ring position 7''$<R<$14'' (see also [58]) using the H$\alpha$ data at its original resolution  which is about the maximum size of the {GMAs observed in CO(2-1) line emission with the Submillimeter Array [24]. }

{Table~1 shows the GMAs' offset position from the galaxy center, $\delta$R.A. and $\delta$Dec. ($\alpha_{2000}=02^{\rm h}\,46^{\rm m}\,18^{\rm s}.96$ and $\delta_{2000}=-30^{\circ}\,16'\,28''.897$), their diameter $d$, the SFR surface density $\Sigma_{\rm SFR}$ as measured by [24], the molecular gas mass surface density $\Sigma_{\rm H_2}$ (adopted for X$_{\rm CO}=1\times 10^{20}$\,cm$^{-2}$\,(K\,km\,s$^{-1}$)$^{-1}$), the turbulent velocity $\sigma_v=\delta$V$_{\rm int}/\sqrt{8\,{\rm ln2}}$ with $\delta$V$_{\rm int}$ the intrinsic CO(2-1) line width given by [24], as well as the B and $<n_e>$ measurements averaged over the GMAs. Also shown are $\beta=E_{\rm th}/E_{\rm b}$ and the mass-to-magnetic flux ratio $\mu$ in units of $\mu_0=(2\pi\,G^{1/2})^{-1}$ (see Sect. 2.2). 

To investigate the efficiency of the GMAs in forming massive stars, we used the {\it star formation per free-fall}, SFR$_{\rm ff}\equiv \Sigma_{\rm SFR}/\Sigma_{\rm H_2} \times t_{\rm ff}$  that is a dimensionless efficiency, SFR$_{\rm ff} =$ SFE $\times t_{\rm ff}$ with $t_{\rm ff}= 4.7 \left(\frac{M}{10^6\,M_{\odot}}\right)^{0.25}$\,Myr [59] with $M=M_{\rm H_2}$ given by [24] and adopted for X$_{\rm CO}=1\times 10^{20}$\,cm$^{-2}$\,(K\,km\,s$^{-1}$)$^{-1}$. Figure~\ref{fig:SFE} also compares the molecular gas depletion timescale $\tau_{\rm d}\equiv\,$1/SFE$=\Sigma_{\rm H_2}/\Sigma_{\rm SFR}$ of the GMAs. 

Through the paper, the errors show both statistical and systematic uncertainties. }

\paragraph{Data Availability}
Source data for figures 3 and 4 are provided in Table~1. The thermal and nonthermal RC data are available from the corresponding author upon reasonable request.

\subsubsection*{References}
{\small

\begin{itemize}
\item[1] Schaye, J. et al. The EAGLE project: simulating the evolution and assembly of galaxies and their environments. {\it Mon. Not. R. Astron. Soc.} {\bf 446}, 521-554 (2015).\\

\item[2] Gatto, A. et al. The SILCC project - III. Regulation of star formation and outflows by stellar winds and supernovae. {\it Mon. Not. R. Astron. Soc.} {\bf 466}, 1903-1924 (2017).\\

\item[3] Gressel, O. et al. Direct simulations of a supernova-driven galactic dynamo. {\it Astron. Astrophys.} {\bf 486}, L35-L38 (2008).\\

\item[4] Xu, H. et al. Turbulence and Dynamo in Galaxy Cluster Medium: Implications on the Origin of Cluster Magnetic Fields. {\it Astrophys. J.} {\bf 698}, L14-L17 (2009).\\

\item[5] Wareing, C.~J. et al. Magnetohydrodynamic simulations of mechanical stellar feedback in a sheet-like molecular cloud. {\it Mon. Not. R. Astron. Soc.} {\bf 465}, 2757-2783 (2017).\\

\item[6] Beck, R. Magnetism in the spiral galaxy NGC 6946: magnetic arms, depolarization rings, dynamo modes, and helical fields. {\it Astron. Astrophys.} {\bf 470}, 539-556 (2007).\\

\item[7] Tabatabaei, F.~S., et al. High-resolution radio continuum survey of M 33. III. Magnetic fields. {\it Astron. Astrophys.} {\bf 490}, 1005-1017 (2008).\\

\item[8] Yoast-Hull, T. M., et al. Winds, Clumps, and Interacting Cosmic Rays in M82. {\it Astrophys. J.} {\bf 768}, 53-68 (2013).\\

\item[9] Tabatabaei, F.~S., et al. The Radio Spectral Energy Distribution and Star-formation Rate Calibration in Galaxies. {\it Astrophys. J.} {\bf 836}, 185-209 (2017).\\

\item[10] Tabatabaei, F.~S. Uncovering star formation feedback and magnetism in galaxies with radio continuum surveys. {\it Highlights on Spanish Astrophysics} {\bf IX}, AUTHOR ISBN 978-84-606-8760-3 (2017).\\

\item[11] V{\'a}zquez-Semadeni, E. et al. Molecular cloud evolution - IV. Magnetic fields, ambipolar diffusion and the star formation efficiency. {\it Mon. Not. R. Astron. Soc.} {\bf 414}, 2511-2527 (2011).\\

\item[12] K\"ortgen, B. and Banerjee, R. Impact of magnetic fields on molecular cloud formation and evolution. {\it Mon. Not. R. Astron. Soc.} {\bf 451}, 3340-3353 (2015).\\

\item[13] Crutcher, R.~M. Magnetic Fields in Molecular Clouds. {\it Annu. Rev. Astron. Astrophys.} {\bf 50}, 29-63 (2012). \\

\item[14] Pillai, T. et al. Magnetic Fields in High-mass Infrared Dark Clouds. {\it Astrophys. J.} {\bf 799}, 74-81 (2015). \\

\item[15] Col\'{\i}n, P. et al. Molecular cloud evolution - V. Cloud destruction by stellar feedback. {\it Mon. Not. R. Astron. Soc.} {\bf 435}, 1701-1714 (2013). \\

\item[16] Tasker, E. J., et al.  Star Formation in Disk Galaxies. III. Does Stellar Feedback Result in Cloud Death? {\it Astrophys. J.} {\bf 801}, 33-47 (2015).\\

\item[17] Bell, E. F. Galaxy Bulges and their Black Holes: a Requirement for the Quenching of Star Formation. {\it Astrophys. J.} {\bf 682}, 355-360 (2008).\\

\item[18] Tacchella, S. et al. Evidence for mature bulges and an inside-out quenching phase 3 billion years after the Big Bang. {\it Science} {\bf 348}, 314-317 (2015)\\

\item[19] Storchi-Bergmann, T. et al. Evidence of a Starburst within 9 Parsecs of the Active Nucleus of NGC 1097. {\it Astrophys. J.} {\bf 624}, L13-L16 (2005).\\

\item[20] Tully, R. B. Nearby galaxies catalog. (Cambridge University Press, 1988).\\

\item[21] Higdon, J.~L. and Wallin, J.~F., A Minor-Merger Interpretation for NGC 1097's ``Jets''. {\it Astrophys. J.} {\bf 585}, 281-297 (2003). \\

\item[22] Martin, D.~C. et al. The UV-Optical Galaxy Color-Magnitude Diagram. III. Constraints on Evolution from the Blue to the Red Sequence. {\it Astrophys. J. Suppl. Ser. } {\bf 173}, 342-356 (2007).\\

\item[23] Salim, S. et al. UV Star Formation Rates in the Local Universe. {\it Astrophys. J. Suppl. Ser. } {\bf 173}, 267-292 (2007).\\

\item[24] Hsieh, P. Y. et al. Physical Properties of the Circumnuclear Starburst Ring in the Barred Galaxy NGC 1097. {\it Astrophys. J.} {\bf 736}, 129-146 (2011).\\

\item[25] {Mart{\'{\i}}n}, S. et al. Multimolecule ALMA observations toward the Seyfert 1 galaxy NGC 1097. {\it Astron. Astrophys.} {\bf 573}, 116-129 (2015). \\

\item[26] Prieto, M. A. et al. Feeding the Monster: The Nucleus of NGC 1097 at Subarcsecond Scales in the Infrared with the Very Large Telescope. {\it Astron. J.} {\bf 130}, 1472-1481 (2005).\\

\item[27] Beck, R. et al. Magnetic fields in barred galaxies. IV. NGC 1097 and NGC 1365. {\it Astron. Astrophys.} {\bf 444}, 739-765 (2005). \\

\item[28] Jogee, S. et al. The Central Region of Barred Galaxies: Molecular Environment, Starbursts, and Secular Evolution. {\it Astrophys. J.} {\bf 630}, 837-863 (2005). \\

\item[29] {Mezcua}, M. and {Prieto}, M.~A. Evidence of Parsec-scale Jets in Low-luminosity Active Galactic Nuclei. {\it Astrophys. J.} {\bf 787}, 62-72 (2014). \\

\item[30] {Fern{\'a}ndez-Ontiveros}, J.~A. et al. Far-infrared Line Spectra of Active Galaxies from the Herschel/PACS Spectrometer: The Complete Database. {\it Astrophys. J. Suppl. Ser. } {\bf 226}, 19-45 (2016). \\

\item[31] Gaensler, B. M. et al.  The Vertical Structure of Warm Ionised Gas in the Milky Way. {\it Publ. Astron. Soc. Aust.} {\bf 25}, 184-200 (2008). \\

\item[32] Tabatabaei, F. S. et al. A detailed study of the radio-FIR correlation in NGC 6946 with Herschel-PACS/SPIRE from KINGFISH. {\it Astron. Astrophys.} {\bf 552}, 19-37 (2013). \\

\item[33] {Kazantsev}, A.~P. Enhancement of a Magnetic Field by a Conducting Fluid. {\it Sov. J. Exp. Theor. Phys.} {\bf 26}, 1031 (1968). \\

\item[34] {Schleicher}, D.~R.~G. and {Beck}, R. A new interpretation of the far-infrared - radio correlation and the expected breakdown at high redshift. {\it Astron. Astrophys.} {\bf 556}, 142-154 (2013). \\

\item[35] Tabatabaei, F. S. et al. An Empirical Relation between the Large-scale Magnetic Field and the Dynamical Mass in Galaxies. {\it Astrophys. J.} {\bf 818}, L10-L16 (2016). \\

\item[36] Federrath, C. et al.  A New Jeans Resolution Criterion for (M)HD Simulations of Self-gravitating Gas: Application to Magnetic Field Amplification by Gravity-driven Turbulence. {\it Astrophys. J.} {\bf 731}, 62-78 (2011). \\

\item[37] {Ferri{\`e}re}, K.~M. The interstellar environment of our galaxy. {\it Rev. Mod. Phys.} {\bf 73}, 1031-1066 (2001). \\

\item[38] {Ondrechen}, M.~P. et al. H I in barred spiral galaxies. II - NGC 1097. {\it Astrophys. J.} {\bf 342}, 39-48 (1989). \\

\item[39] Tabatabaei, F. S. et al. Cold dust in the giant barred galaxy NGC 1365. {\it Astron. Astrophys.} {\bf 555}, 128-139 (2013). \\

\item[40] Murgia M. et al. The molecular connection to the FIR-radio continuum correlation in galaxies. {\it Astron. Astrophys.} {\bf 437}, 389-410 (2005).\\

\item[41]  Yoast-Hull, T. M., Gallagher III, J. S., Zweibel, E. G. Equipartition and Cosmic Ray Energy Densities in Central Molecular Zones of Starbursts. {\it Mon. Not. R. Astron. Soc.} {\bf 457}, L29-L33 (2016). \\

\item[42] {Nakano}, T. and {Nakamura}, T. Gravitational Instability of Magnetized Gaseous Disks 6. {\it Publ. Astron. Soc. Jpn.} {\bf 30}, 671-680 (1978). \\

\item[43] Basu, S., Mouschovias, T. Ch. Magnetic Braking, Ambipolar Diffusion, and the Formation of Cloud Cores and Protostars. III. Effect of the Initial Mass-to-Flux Ratio. {\it Astrophys. J.} {\bf 453}, 271-283 (1995). \\

\item[44] {Heiles}, C. and {Crutcher}, R. Magnetic Fields in Diffuse HI and Molecular Clouds in 'Cosmic Magnetic Fields'. {\it Lect. Notes Phys.} {\bf 664}, 137 (2005). \\

\item[45] Allard, E. L., Knapen, J. H., Peletier, R. F., \& Sarzi, M. The star formation history and evolution of the circumnuclear region of M100. {\it Mon. Not. R. Astron. Soc.} {\bf 371}, 1087-1105 (2006).\\

\item[46] Mouschovias, T.~C. Magnetic braking, ambipolar diffusion, cloud cores, and star formation - Natural length scales and protostellar masses. {\it Astrophys. J.} {\bf 373}, 169-186 (1991). \\

\item[47] Li, H. and Henning, T. The alignment of molecular cloud magnetic fields with the spiral arms in M33. {\it Nature} {\bf 479}, 499-501 (2011). \\

\item[48] {Krumholz}, M.~R. and {McKee}, C.~F. A minimum column density of 1gcm$^{-2}$ for massive star formation. {\it Nature} {\bf 451}, 1082-1084 (2008). \\

\item[49] Fletcher, A. et al. Magnetic fields and spiral arms in the galaxy M51. {\it Mon. Not. R. Astron. Soc.} {\bf 412}, 2396-2416 (2011).\\

\item[50] Tabatabaei, F. S. et al. Multi-scale radio-infrared correlations in M 31 and M 33: The role of magnetic fields and star formation. {\it Astron. Astrophys.} {\bf 557}, 129-143 (2013). 

\item[51] Tabatabaei, F. S. et al. High resolution radio continuum survey of M33: II. Thermal and nonthermal emission. {\it Astron. Astrophys.} {\bf 475}, 133-143 (2007). \\

\item[52] Mezcua, M. et al. The warm molecular gas and dust of Seyfert galaxies: two different phases of accretion? {\it Mon. Not. R. Astron. Soc.} {\bf 452}, 4128-4144 (2015). \\

\item[53] Phillips, M. M. et al. Nuclear activity in two spiral galaxies with jets: NGC 1097 and 1598. {\it Mon. Not. R. Astron. Soc.} {\bf 210}, 701-710 (1984). \\

\item[54] Calzetti, D. et al. The Dust Content and Opacity of Actively Star-forming Galaxies. {\it Astrophys. J.} {\bf 533}, 682-695 (2000). \\

\item[55] Osterbrock, D. E. Astrophysics of gaseous nebulae and active galactic nuclei. (University Science Books, 1989). \\

\item[56] Ehle, M. and Beck. R. Ionized Gas and Intrinsic Magnetic Fields in the Spiral Galaxy NGC6946. {\it Astron. Astrophys.} {\bf 273}, 45-64 (1993). \\

\item[57] Beck, R. and {Krause}, M. Revised equipartition and minimum energy formula for magnetic field strength estimates from radio synchrotron observations. Astronomische Nachrichten {\bf 326}, 414-427 (2005). \\

\item[58] Comer\'on, S. et al. AINUR: Atlas of Images of NUclear Rings. {\it Mon. Not. R. Astron. Soc.} {\bf 402}, 2462-2490 (2010).\\

\item[59] Krumholz, M.~R. and McKee, C.~F. A General Theory of Turbulence-regulated Star Formation, from Spirals to Ultraluminous Infrared Galaxies. {\it Astrophys. J.} {\bf 630}, 250-268 (2005).
\end{itemize}
}

\end{document}